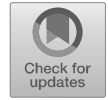

# Data based constitutive modelling of rate independent inelastic effects in composite cables using Preisach hysteresis operators

Davide Manfredo[1] · Vanessa Dörlich[1] · Joachim Linn[1] · Martin Arnold[2]

**Abstract**
This contribution aims at introducing first steps to develop hysteresis operator type inelastic constitutive laws for Cosserat rods for the simulation of cables composed of complex interior components. Motivated by the basic elements of Cosserat rod theory, we develop a specific approach to constitutive modelling adapted for this application. Afterwards, we describe the hysteretical behaviour arising from cyclic bending experiments on cables by means of the Preisach operator. As shown in pure bending experiments, slender structures such as electric cables behave inelastically, and open hysteresis loops arise with noticeable difference between the first load cycle and the following ones. The Preisach operator plays an important role in describing the input-output relation in hysteresis behaviours, and it can be expressed as a superposition of relay operators. Hence, a mathematical formulation of the problem is introduced, and a first attempt is made to determine the hysteresis behaviour that describes the relation between curvature and bending moment. Therefore, a suitable kernel function is identified in a way that its integration over the Preisach plane results in the bending moment of the specimen, and a comparison between different kernel functions is performed.

## 1 Introduction

Electric cables, as those shown in Fig. 1, are complex objects due to their multi-material composition and their geometric properties [8, 9]. Consequently, different internal interaction effects occur and lead to an observed effectively inelastic deformation behaviour of such

✉ D. Manfredo
davide.manfredo@itwm.fraunhofer.de

J. Linn
joachim.linn@itwm.fraunhofer.de

1  Fraunhofer ITWM, Fraunhofer Platz 1, 67663 Kaiserslautern, Germany

2  Institute of Mathematics, Martin Luther University Halle-Wittenberg, Theodor-Lieser-Str. 5, 06120 Halle (Saale), Germany

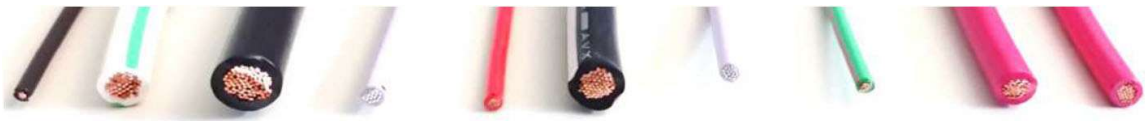

**Fig. 1** Cross sections of different electric cables. (Colour figure online)

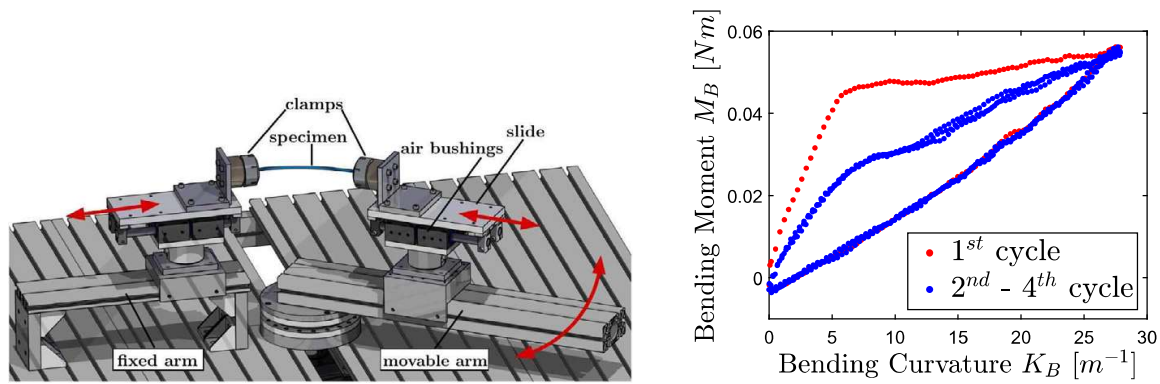

**Fig. 2** *Left*: Pure bending test rig. *Right*: Bending moment vs. bending curvature diagram measured in a pure bending experiment [9]. (Colour figure online)

cables. Cyclic bending experiments show open hysteresis loops with noticeable difference between the first load cycle and the following ones, as shown in Fig. 2 *(right)*. In this regard, efforts have been made to develop effective inelastic constitutive models within the framework of geometrically exact Cosserat rod theory [7]. A similar aim was pursued by [27, 28], where inelastic constitutive behaviour was considered in the modelling and simulation of manufactured composite materials, also described in terms of 3D Cosserat rods.

Figure 2 *(left)* describes the setup of the pure bending experiment as introduced in [8], which enables direct access to the bending moment and bending curvature. By applying only a bending moment on the specimen, a deformation state of pure bending is achieved. The configuration of the experiment ensures that no normal or shear forces act on the specimen during the experiment. The centreline of the specimen is bent into a circular arc with constant bending curvature $K_B$ and consequently constant bending moment $M_B$ along the specimen during the test. Therefore, this experimental setup yields the possibility to directly observe the effective local constitutive behaviour of the measured specimen, given in terms of the time history of value pairs $(M_B(s,t), K_B(s,t))$ of the bending moment and associated curvature, for all times $t$ of the recorded measurement process, where for each $t$ the values at different centreline positions $s \in [0, L]$ are the same due to the special experimental setup.

The essential components of a continuum mechanical model [2, 10, 19] used to simulate (isothermal) spatial deformations of flexible solid bodies are:

– a geometric model that describes the *kinematics* in terms of configuration variables of the spatial position of the material points, as well as differential *strain* measures that capture local material deformations w.r.t. a (stress free) reference configuration,
– the governing *balance equations* that express the local static or dynamic equilibrium of all forces acting within the body in terms of a PDE to be satisfied by the *stress* tensor field,
– *boundary conditions* that model prescribed positions (or motions) of parts of the body, as well as given forces acting on its boundary, and

– a *constitutive law* describing the functional relation between the *local stresses* at a certain point of time $t$ and the *time history* of the respective *local strains* that have occurred during the deformation process up to this time.

A Cosserat rod [2, 23] is a kinematically reduced model that describes the shape and spatially deformed configurations of a slender flexible body in terms of a *framed curve*, as sketched in Fig. 3. Based on this kinematical ansatz, the balance equations are expressed in terms of cross section integrated forces and moments. According to [23], these balance equations can be derived from those of the three-dimensional theory in a straightforward manner, and hold like the latter without any assumptions on the constitutive behaviour. The same kind of transfer can be achieved for the boundary conditions. We will provide additional information on this topic in Sect. 2.

To complete the setup of a Cosserat rod model and close the system of governing equations, one needs to formulate corresponding constitutive laws that relate the strain measures of a Cosserat rod to its sectional forces and moments. In the case of an elastic structural behaviour, this can be achieved via an elastic potential energy function depending on the strains with constitutive equations derived by taking the gradient of the potential w.r.t. the strains.

The task becomes considerably more demanding in the case of *inelastic* structural behaviour, which is of primary interest here. Besides general theoretical considerations as discussed in the monographs of Antman [2] (see Chap. 12, Sect. 10), Haupt [10] (see Chap. 7) or Truesdell and Noll [25] (see Chap. C, Sect. III), the work of Simo and Hughes [24] provides a comprehensive presentation of a state of the art approach to elastoplastic, viscoplastic and viscoelastic constitutive models of rheological type, defined in terms of evolution equations of internal inelastic stress or strain variables. The recent article of Bauchau and Nemani [4] utilises a selection of such rheological constitutive models to handle inelastic effects in structural components of flexible multibody systems, including nonlinear beams.

Such rheological constitutive models are formulated in terms of ODEs or DAEs with a priori defined r.h.s. functions of more or less simple mathematical form, usually depending on a few model parameters. While the recent work of Dörlich et al. [7] shows that a constitutive model of this type can be successfully applied to reproduce the observed bending behaviour of a single composite cable, one cannot expect that the same can be achieved for other cables of different internal structure—note the variety displayed in Fig. 1—by merely adapting the parameters of the same ODE or DAE model.

This motivated the authors of this work to consider a more flexible, data based approach to formulate constitutive laws for composite cable structures. This finally led to the idea to utilise so called *hysteresis operators* [5, 14, 18, 26] as grey box models. The functional form and building blocks of hysteresis operators capture inelastic effects in a generic, phenomenological manner without the need of a priori assumptions on the material behaviour.

In this work, the authors present first steps on the path of their research efforts to develop hysteresis operator type constitutive laws for Cosserat rods, applied to simulate large deformations of composite cables. The article is structured as follows: In Sect. 2 we provide a compact summary of Cosserat rod theory basics, comprising an approach to constitutive modelling where Cosserat rods are utilised as grey box models for cable simulation. We proceed with an introduction of hysteresis operators in Sect. 3 with particular focus on the relay and the Preisach operator. Section 4 is dedicated to the mathematical formulation of the problem of approximating the bending moment vs. bending curvature diagram by means of the Preisach operator, while Sect. 5 concentrates on the comparison of different kernel functions identified starting from different experimental data.

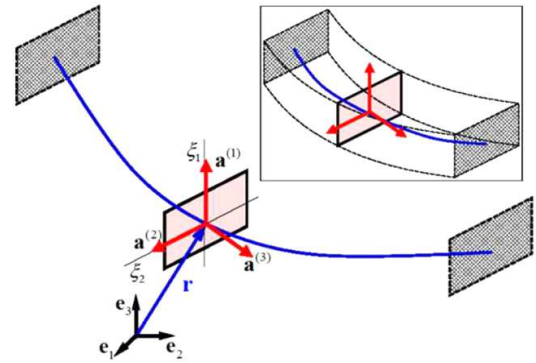

**Fig. 3** Spatial configuration of a Cosserat rod, represented by a centreline curve $\mathbf{r}(s, t)$ with attached moving frame $\mathsf{R}(s,t) = \mathbf{a}^{(k)}(s,t) \otimes \mathbf{e}_k$ (see [16] for details). (Colour figure online)

## 2 Cosserat rods as grey box models for cable simulation

The European project THREAD[1] investigates novel approaches to model and simulate slender, highly flexible beam-like structures for a broad variety of industrial applications, e.g. in automotive, textile and medical industry [3].

Cables as those shown in Fig. 1 are slender flexible objects which in practical applications are deformed in space mainly by bending. Examples are the manufacturing of wiring harnesses and their assembly during the production process of passenger cars, commercial vehicles, airplanes, ships and boats, or other complex mechatronic machinery [17]. Although any twisting of cables should be avoided to prevent damage, it is mainly the *complex interplay between bending and torsion* that governs the shape of spatially deformed cables, so any reasonable simulation model must account for these deformation modes. This implies that we need to reach beyond the notion of a "flexible curve", which leads us more or less directly to the theory of *geometrically exact rods* as a proper modelling framework for cable simulation.

The theory of Cosserat rods [1, 2, 21, 23] provides a consistent framework to model large spatial deformations of slender flexible beam-like structures by bending and twisting. As sketched in Fig. 3, the configuration of a Cosserat rod is described by a *framed curve*.

For the purpose of cable simulation, we need to have a closer look on the kinematic assumptions. At first, a cable is not a homogeneous connected body, but a multitude of (in some cases: extremely) slender components that are by no means in perfect contact, but interact mutually in a manner that can practically neither be observed nor simulated in detail. From a macroscopic point of view that monitors cable deformations on a gross level, the framed curve model possesses the necessary degrees of freedom to model these deformations. Therefore we suggest to stick with the kinematics of the framed curve model for cable simulation applications, combined with the working hypothesis that the effects of internal structural interactions can be captured at a sufficient level of accuracy by data based constitutive modelling (see Sect. 2.1 below).

**Strain measures of Cosserat rods** As discussed in detail in [16], the *differential invariants* of a framed curve given by its material tangent vector $\mathbf{\Gamma} = \mathsf{R}^T \cdot \partial_s \mathbf{r}$ and material curvature vector $\mathbf{K} = \mathsf{R}^T \cdot \boldsymbol{\kappa}$ determine its geometry in Euclidean space uniquely up to global rigid body motions and therefore serve as frame indifferent *strain measures* of a Cosserat rod, where $\mathbf{K}$ and $\mathbf{\Gamma}$ contain the components of the spatial curvature vector $\boldsymbol{\kappa}$, implicitly given by the generalised Frenet equation $\partial_s \mathbf{a}^{(j)} = \boldsymbol{\kappa} \times \mathbf{a}^{(j)}$, and the components of the centreline tangent vector $\partial_s \mathbf{r}$ w.r.t. the local frame.

---

[1]THREAD (thread-etn.eu) – Joint training on numerical modelling of highly flexible structures for industrial applications.

The deviations $\mathbf{\Gamma}(s,t) - \mathbf{\Gamma}_0$ and $\mathbf{K}(s,t) - \mathbf{K}_0(s)$ of the invariants from their reference values $\mathbf{\Gamma}_0 = \mathbf{e_3}$ and $\mathbf{K}_0(s)$ measure the local deformation of a Cosserat rod in a frame indifferent way, where $\mathbf{\Gamma}_0 = \mathbf{e_3} \Leftrightarrow \partial_s \mathbf{r}_0(s) = \mathbf{a}^{(3)}(s)$ implies that in the reference configuration the centreline curve is parameterised by arc length, and the cross sections are oriented orthogonal to the centreline.

Consistent with our working hypothesis that the framed curve kinematics of a Cosserat rod provide a proper description of the configuration geometry for deformed cables, we intend to utilise the invariants $\mathbf{\Gamma}$ and $\mathbf{K}$, or more precisely their time histories, as input variables for the formulation of inelastic constitutive laws in Sect. 2.1.

Different from other applications of nonlinear rod models in structural mechanics, Cosserat rods as simulation models for composite cables have the nature of *grey box* models due to limitations implied by the fact that the complex internal structure of composite cables and details of the mutual interaction of their constituents are neglected.

## 2.1 Elastic and inelastic constitutive models for Cosserat rods

The formulation of inelastic constitutive laws suitable to describe the deformation of composite cables is the third building block of our modelling framework. In 3D solid mechanics [2, 10], one would start in an abstract manner with a functional relation like

$$\mathsf{P}(\mathbf{x},t) = \mathsf{F}(\mathbf{x},t) \cdot \mathsf{S}(\mathbf{x},t) \,, \ \mathsf{S}(\mathbf{x},t) = \hat{\mathcal{S}}[\mathsf{C}(\mathbf{x},t)] \,.$$

This describes the general form of an *elastic* constitutive law for the first Piola–Kirchhoff stress $\mathsf{P}$ in terms of a material constitutive function $\hat{\mathcal{S}}[\ldots]$ that yields values of the second Piola–Kirchhoff stress $\mathsf{S}$ for local material strains given in terms of the right Cauchy–Green tensor $\mathsf{C} = \mathsf{F}^T \cdot \mathsf{F}$. Stated in this form, the elastic constitutive law automatically possesses the important fundamental properties of local action (specifically for so called simple materials) and frame indifference (a.k.a. material objectivity).

*Inelastic* constitutive laws with these fundamental properties are then formulated in the more general (abstract) form

$$\mathsf{P}(\mathbf{x},t) = \mathsf{F}(\mathbf{x},t) \cdot \mathsf{S}(\mathbf{x},t) \,, \ \mathsf{S}(\mathbf{x},t) = \hat{\mathcal{S}}[\mathsf{C}(\mathbf{x},\tau)]^t \,, \tau \leq t \,,$$

where the notation $\hat{\mathcal{S}}[\ldots]^t$ of the functional indicates that its value at the present time $t$ depends on the whole strain history of the deformation process for times $\tau$ before $t$. Applied to our Cosserat modelling framework for cable simulation, we utilise both types of constitutive laws in parts.

For the material forces $\mathbf{F}$, we suggest either to assume an elastic constitutive law in its most simple standard form

$$\mathbf{F}(s,t) = \text{diag}([GA],[GA],[EA]) \cdot [\mathbf{\Gamma}(s,t) - \mathbf{\Gamma}_0] \,, \tag{1}$$

with effective elastic stiffness parameters $[GA]$ and $[EA]$ for transverse shearing and longitudinal extension, or to inhibit one or both of these deformation modes, such that the transverse and longitudinal components of $\mathbf{F}$ become Lagrangian multipliers in the equilibrium equations of an extensible or inextensible Kirchhoff rod model.

Moreover, we consider an inelastic constitutive model of the form

$$\mathbf{M}(s,t) = \hat{\mathcal{M}}[\mathbf{K}(s,\tau)]^t \,, \tau \leq t \tag{2}$$

for the material moments, where the vector-valued functional $\hat{\mathcal{M}}[\ldots]^t$ yields the bending and torsional moment depending on the history of material curvatures $\mathbf{K}(s, \tau)$ for times $\tau$ before and up to $t$.

We may further specialise the abstract constitutive model (2) to scalar form to focus on the case of plane bending, as discussed in the introductory section about the experimental setup:

$$M_B(s,t) = \hat{\mathcal{M}}_B[K_B(s, \tau)]^t, \tau \leq t. \tag{3}$$

We proceed in the following section with the construction of such a constitutive model of *integral* type, using the functional form of a *hysteresis operator*.

## 3 Hysteresis operators

Hysteresis refers to a relation between two scalar time-dependent quantities that cannot be expressed in terms of a single-valued function [5]. The relation describes a process with rate independent memory [26], i.e. the output is invariant with respect to changes of the time scale, and it may not only depend on the value of the input at time $t$, but also on its previous evolution. As shown in [5, 18, 26], hysteresis and hysteresis operators are a well-studied topic with a variety of applications, mainly hysteresis effects arising from electric and magnetic phenomena. The Preisach operator $\mathcal{P}$ plays a major role in modelling the input–output relation in hysteresis behaviours and can be expressed as a superposition of relay operators $\mathcal{R}$.

In this section, we will denote by $v$ any input function and by $w$ any output function, even when talking about specific cases (e.g. bending curvature and bending moment), whereas in the next section the notation will become more specific.

### 3.1 Relay operator

Given any couple $(a_1, a_2) \in \mathbb{R}^2$ with $a_1 < a_2$, we introduce the relay operator $\mathcal{R}_{a_1,a_2}$. For any continuous input function $v \in \mathcal{C}([0, t_E])$, starting from an initial value $\xi \in \{\pm 1\}$, the output

$$w = \mathcal{R}_{a_1,a_2}[v, \xi] : [0, t_E] \to \{\pm 1\} \tag{4}$$

will be equal to $-1$ if the input function value $v(t)$ crosses the threshold value $a_1$ from above, and will be equal to $+1$ if $v(t)$ crosses the threshold value $a_2$ from below. Formally, this can be expressed as

$$w(0) := \begin{cases} -1, & v(0) \leq a_1 \\ \xi, & a_1 < v(0) < a_2 \\ +1, & v(0) \geq a_2 \end{cases} \tag{5}$$

and for any $t \in \,]0, t_E]$, setting $X_t = \{\tau \in \,]0, t] : v(\tau) = a_1 \text{ or } v(\tau) = a_2\}$,

$$w(t) := \begin{cases} w(0), & X_t = \emptyset \\ -1, & X_t \neq \emptyset \text{ and } v(\max(X_t)) = a_1 \\ +1, & X_t \neq \emptyset \text{ and } v(\max(X_t)) = a_2. \end{cases} \tag{6}$$

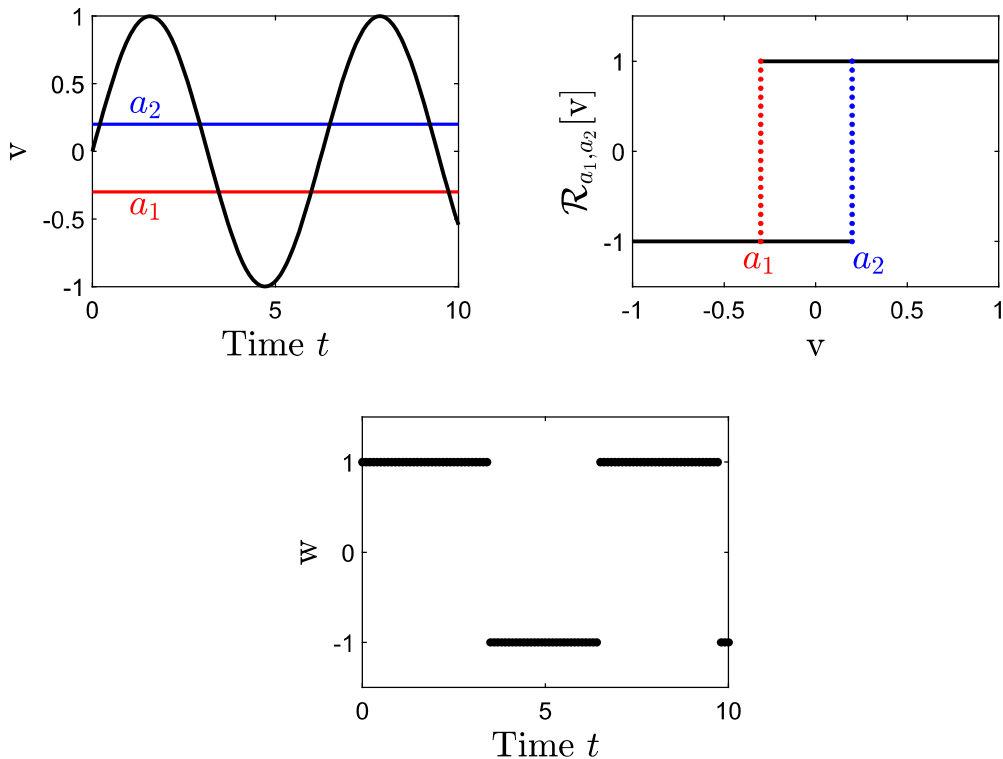

**Fig. 4** *Top left:* input function $v(t) = \sin(t)$ with $t \in [0, 10]$. *Top right:* diagram of the relay operator with $a_1 = -0.3$ and $a_2 = 0.2$. *Bottom:* output function $w(t) = \mathcal{R}_{a_1,a_2}[v,\xi](t)$ with initial value $\xi = +1$. (Colour figure online)

The relay operator can be interpreted as a switch operator between the values $-1$ and $+1$ with switching interval of width $a_2 - a_1$ and centred in $\frac{a_2+a_1}{2}$. A graphical representation of the relay operator is given in Fig. 4.

From the next subsection, for simplicity of notation, we will omit the initial value $\xi$ when writing the relay operator $\mathcal{R}_{a_1,a_2}[v](t)$. This choice is justified by the fact that one could assume $\xi = -1$ or $\xi = 1$ without loss of generality.

### 3.2 Preisach operator

In the early 1930s, Preisach investigated whether the magnetisation in ferromagnetic materials adjusts without inertia to the applied magnetic field [20]. Performing switching experiments allowed him to affirm the question, and he postulated that the magnetisation depends on the magnetic field through a linear superposition of relay operators. Hence, the previously described relay operator is the "building block" of the Preisach operator, which is in fact defined as a superposition of relay operators multiplied by a suitable kernel function $\omega(r, s)$, assumed to vanish for large values of $|s|$ and $r$,

$$w(t) = \mathcal{P}[v](t) = \int_0^{+\infty} \int_{-\infty}^{+\infty} \omega(r,s) \mathcal{R}_{s-r,s+r}[v](t) \, ds \, dr. \qquad (7)$$

Here, $v(t)$ and $w(t)$ are respectively the input (Fig. 5 *top left*) and the output function, $s$ and $r$ are the coordinates of the Preisach plane, and $\mathcal{R}_{s-r,s+r}[v](t) \in \{\pm 1\}$ is the relay operator with switching interval of width $2r$ and centred in $s$. Preisach provided a simple geometrical interpretation for the operator $\mathcal{P}$, which turns out to be very useful from an operative point

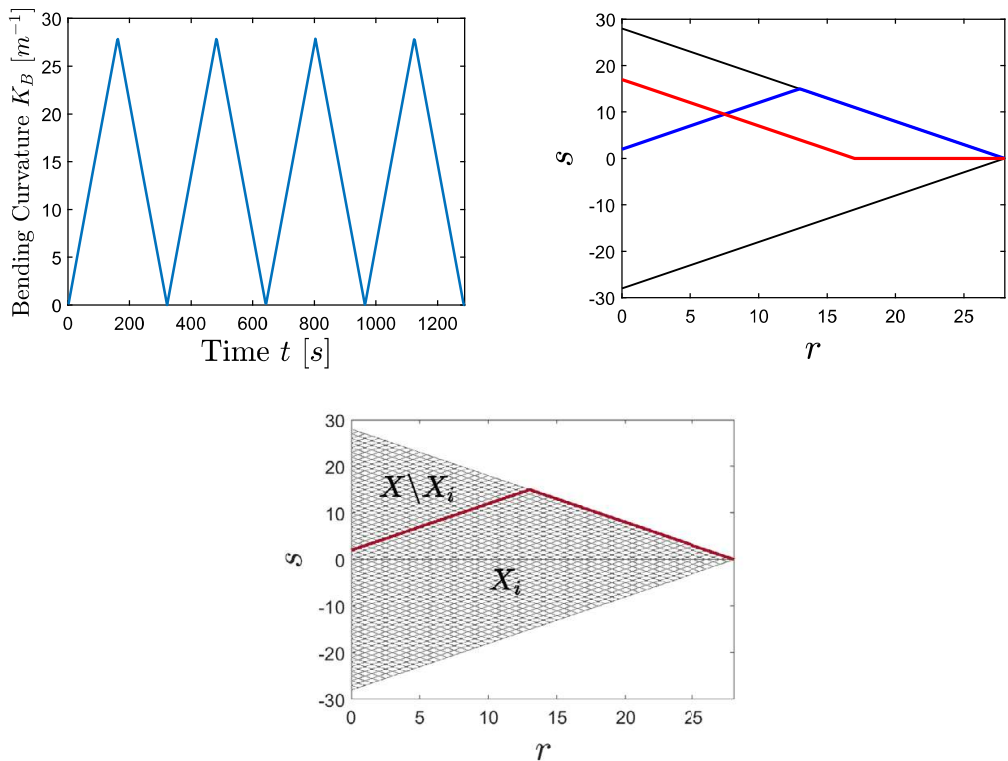

**Fig. 5** *Top left*: input given as curvature vs. time. *Top right*: domain (black triangle) included in the Preisach plane with two examples of memory curve. *Bottom*: domain included in the Preisach plane with an example of triangulation and a memory curve. (Colour figure online)

of view. If we consider an input function $v(t)$, for every time $t$ we determine the sets

$$A_{\pm}(t) = \{(r,s) \in \mathbb{R}_+ \times \mathbb{R} : \mathcal{R}_{s-r,s+r}[v](t) = \pm 1\}. \tag{8}$$

The union of such sets corresponds to the so-called *Preisach plane*.

Let $v : [0, t_E] \to \mathbb{R}$ be a piecewise monotone function with monotonicity partition $0 = t_0 < t_1 < \cdots < t_N = t_E$. One can verify (see [5]) that the dividing line

$$B(t) = \partial A_+(t) \cap \partial A_-(t), \tag{9}$$

also called memory curve, at each time $t$ is the graph of a function which can be defined recursively by

$$\begin{aligned} \psi(0) &= \max\{g^-(v(0)), \min\{g^+(v(0)), 0\}\}, \\ \psi(t) &= \max\{g^-(v(t)), \min\{g^+(v(t)), \psi(t_i)\}\} \\ &\quad \text{for } t_i < t \leq t_{i+1}, \quad 0 \leq i \leq N-1. \end{aligned} \tag{10}$$

Here, $g^-(v)$ denotes the straight line with slope $-1$ through the point $(0, v)$ and $g^+(v)$ denotes the straight line with slope $+1$ through the point $(0, v)$. Note that $B(t)$ carries the total memory information present in the system at time $t$ (see [5]).

In Fig. 5, starting from a specific input function (Fig. 5 *top left*) two examples of memory curves are deduced and depicted in Fig. 5 (*top right*). Figure 5 (*top right*) depicts also one specific choice of subset of Preisach plane, i.e. the triangle $\{(r, s) | r \in [0, 28], -28 + r \leq s \leq 28 - r\}$. If we consider the blue line to be the memory curve $B(t_\alpha)$ for some $t_\alpha$, the sets $A_+(t_\alpha)$ and $A_-(t_\alpha)$ will be the parts of the Preisach plane respectively below and above the blue line. Analogously, assuming the red line to be the memory curve $B(t_\beta)$ for some $t_\beta$, $A_+(t_\beta)$ and $A_-(t_\beta)$ are the subsets of the Preisach plane respectively below and above the red line.

One can observe how the description of the Preisach plane depends on the specific input that is being considered as it will be explained more rigorously in the next section. In fact, in Fig. 5 (*top right*), the set $\{(r,s)|r \in [0, 28], -28 + r \leq s \leq 28 - r\}$ is determined by

$$\max_{t \in [0,t_E]} v(t) = 28 \text{ and } \min_{t \in [0,t_E]} v(t) = 0, \tag{11}$$

where $v : [0, t_E] \to \mathbb{R}$ is the input function depicted in Fig. 5 (*top left*).

From this, one can already understand that the choice of a suitable subset of the Preisach plane could be modified depending on the features of the input function, as well as other specifics of the studied problem. For example, the most common choice is to work over a triangular subset of the Preisach plane, but one could also choose a rectangular subset (see [5]).

Using $\mathcal{R}_{s-r,s+r}[v](t) \in \{\pm 1\}$ and the definition of $A_\pm(t)$, (4) can be rewritten as

$$w(t) = \iint_{A_+(t)} \omega(r,s) ds dr - \iint_{A_-(t)} \omega(r,s) \, ds dr. \tag{12}$$

It should be noted that Preisach hysteresis operators provide a model for causal response (see [26]), such that the output value $w(t)$ at time $t$ depends only on inputs $v(\tau)$ at past times $\tau \leq t$. Thus, hysteresis loops can be computed by integrating a suitable kernel function $\omega(r, s)$ over a domain included in the Preisach plane.

## 4 Approximation problem

As previously mentioned, we would like to describe the relation input–output (bending curvature–bending moment) by means of the Preisach operator, utilising data coming from a pure bending cyclic experiment. This translates into finding a suitable kernel function $\omega(r, s)$ such that its integral over the Preisach plane results in a good approximation of the measured output. This topic has been of interest in the past decades, and several approaches have been presented, such as approximation techniques based on a finite number of optimally chosen experiments [12], approximation by means of neural networks [6, 29] and approximation by means of least square methods [11, 13, 22]. In this work, we utilise the latter.

### 4.1 Problem formulation

As shown in Fig. 2 (*right*), we deal with measurements of bending moment values with respect to bending curvature value during four consecutive load cycles. The dataset that we will consider consists of $\{t_i\}_{i=1}^T$, bending curvature $\{K_{B_i}\}_{i=1}^T$ and bending moment $\{M_{B_i}\}_{i=1}^T$. Note that the values of time and bending curvature are prescribed by the experimental procedure, while the values of bending moment are measured. Moreover, it is relevant to underline that here the time represents rather an order parameter than a time variable, as it is normally considered. Since we are dealing with a rate independent process, the time data could be rescaled and shifted without causing any change in our approach.

Starting from the input function, for each time step $t_i$, we recursively define the Preisach plane, i.e. the sets $A_\pm(t_i)$ and the memory curve $B(t_i)$ from (10). Thus, our goal is to find $\omega(r,s)$ such that the following expression is minimised:

$$\frac{1}{T}\sum_{i=1}^{T}\frac{1}{2}\left(M_{B_i} - \iint_{A_+(t)}\omega(r,s)dsdr + \iint_{A_-(t)}\omega(r,s)dsdr\right)^2. \tag{13}$$

To this end, we will take into account only a subset of the Preisach plane, in particular a subset of $\mathbb{R}_{\geq 0} \times \mathbb{R}$ spanned by the memory curve $B(t)$. More specifically, calling $K_B : [0, t_E] \to \mathbb{R}$ the bending curvature function (Fig. 5 *top left*), setting $m = \max\left(\left|\min_{t \in [0,t_E]} K_B(t)\right|, \left|\max_{t \in [0,t_E]} K_B(t)\right|\right)$, we consider the triangle

$$\{(r,s)|r \in [0,m], -m+r \leq s \leq m-r\}. \tag{14}$$

As shown in [13, 22], we choose a tolerance $d$ to round the input values, and we divide the part of the Preisach plane crossed by the memory curve $B(t)$ in $n$ elements, such that at each time step, $B(t_i)$ lays on their edges (see Fig. 5 *bottom*). We denote by $\{e^m\}_{m=1}^n$ the elements of the triangulation, $X \subset \mathbb{N}_+$ the set of indices given to the elements of the triangulation, and we define the following sets:

$$\begin{aligned} X_i &= \{m \in X | e^m \text{ below the memory curve at time } t_i\}, \\ X \setminus X_i &= \{m \in X | e^m \text{ above the memory curve at time } t_i\}. \end{aligned} \tag{15}$$

We observe that $\bigcup_{m \in X_i}\{e^m\}$ and $\bigcup_{m \in X \setminus X_i}\{e^m\}$ are the discrete corresponding sets of $A_+(t_i)$ and $A_-(t_i)$, respectively.

We assume that the kernel function $\omega(r,s)$ is a piecewise constant over each element of the mesh, and we want to approximate the output as

$$M_{B_i} \approx \sum_{m \in X_i}\iint_{e^m}\omega(r,s)dsdr - \sum_{m \in X \setminus X_i}\iint_{e^m}\omega(r,s)dsdr \quad i=1,\ldots,T. \tag{16}$$

Now, for each time step, we define the row vector $\boldsymbol{\Delta}_i = [\delta_i^1, \ldots, \delta_i^n]$, where

$$\delta_i^m = \begin{cases} 1 & \text{if } m \in X_i \\ -1 & \text{if } m \in X \setminus X_i. \end{cases} \tag{17}$$

Calling $x^m = \iint_{e^m}\omega(r,s)dsdr$, we have

$$\Delta = \begin{bmatrix}\boldsymbol{\Delta}_1 \\ \vdots \\ \boldsymbol{\Delta}_T\end{bmatrix} \in \mathbb{R}^{T \times n}, \quad X = \begin{bmatrix}x^1 \\ \vdots \\ x^n\end{bmatrix} \in \mathbb{R}^n, \quad Y \in \begin{bmatrix}M_{B_1} \\ \vdots \\ M_{B_T}\end{bmatrix} \in \mathbb{R}^T. \tag{18}$$

Hence, the function to be minimised is $f(X) = \frac{1}{2}\|\Delta \cdot X - Y\|^2$.

In our case, we deal with insufficient experimental data (more unknowns than observations), hence the rank of the matrix $\Delta$ is $\text{rank}(\Delta) = q < \min\{T, n\}$. We then need to perform a singular value decomposition of the matrix $\Delta^T \Delta = USV^T$, where $S$ is a diagonal matrix with $\text{rank}(S) = q$.

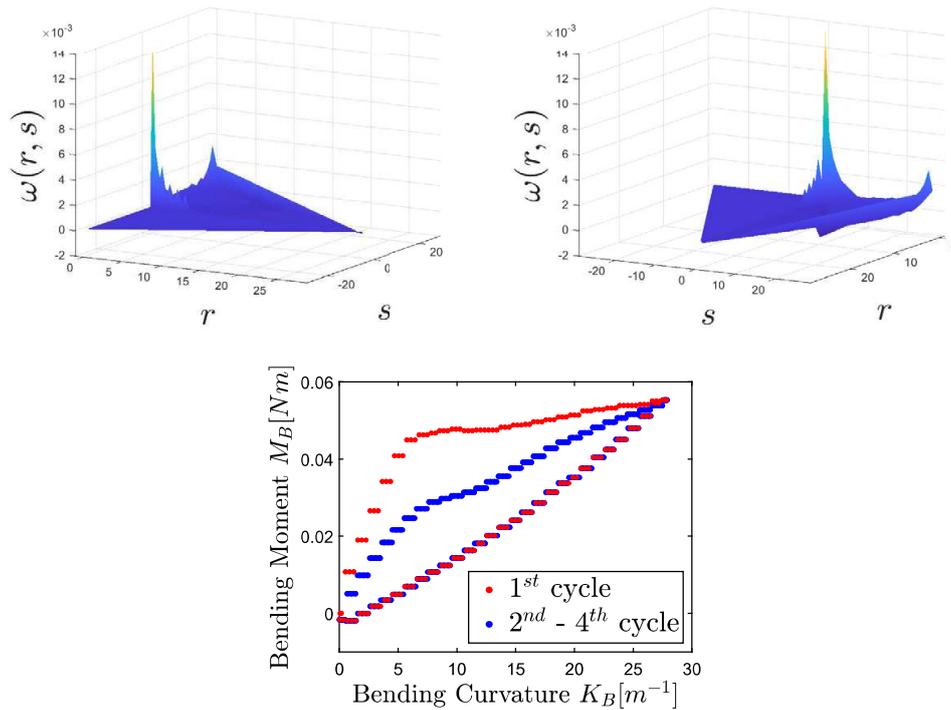

**Fig. 6** *Top left, right:* kernel function obtained by minimising (19), observed by different angles. *Bottom:* estimated plot of bending moment vs. curvature obtained by means of the hysteresis operator. (Colour figure online)

We extract $\hat{S}$, $\hat{U}$, $\hat{V}$ from $S$, $U$, $V$, respectively, by eliminating the rows and the columns of $S$ that are zero and the corresponding columns of $U$ and $V$. Setting $\boldsymbol{X} = \hat{V}\boldsymbol{Z}$, the expression to be minimised becomes

$$g(\boldsymbol{Z}) = \boldsymbol{Z}^T \hat{S} \boldsymbol{Z} - \boldsymbol{Y}^T \Delta \cdot \hat{V} \boldsymbol{Y}. \tag{19}$$

It is easily verified that once a minimiser $\boldsymbol{Z}^*$ of $g$ is found, then $\boldsymbol{X}^* = \hat{V}\boldsymbol{Z}^*$ minimises $f$.

### 4.2 Kernel function and approximated data

A minimiser $\boldsymbol{Z}^*$ of $g$ can be found using the Matlab function *quadprog*. In Fig. 6 (*top left, right*), an approximation of the kernel function $\omega(r, s)$ is shown, and the integral of such a kernel function over the domain included in the Preisach plane results in the diagram shown in Fig. 6 (*bottom*). Comparing the experimental data in Fig. 2 (*right*) with the diagram in Fig. 6 (*bottom*), one can see that this approach describes the relation input–output (i.e. bending curvature–bending moment) observed during the experiments quite well. One should note that the step-like behaviour of the diagram in Fig. 6 (*bottom*) is due to the tolerance value $d$. However, the approximated kernel function shows a highly nonlinear behaviour.

## 5 Comparison of different kernel functions

In this section we apply the described procedure to pure bending data coming from a similar yet different pure bending experiment using the same cable type as in Fig. 2 (*right*). As shown in Fig. 7, the experiment consists of a higher number of load cycles with increasing maximum bending curvature. For each maximum bending curvature $\{K_B^k\}_{k=1}^{9}$, three cycles are performed. Afterwards, the curvature is increased until the next maximum bending

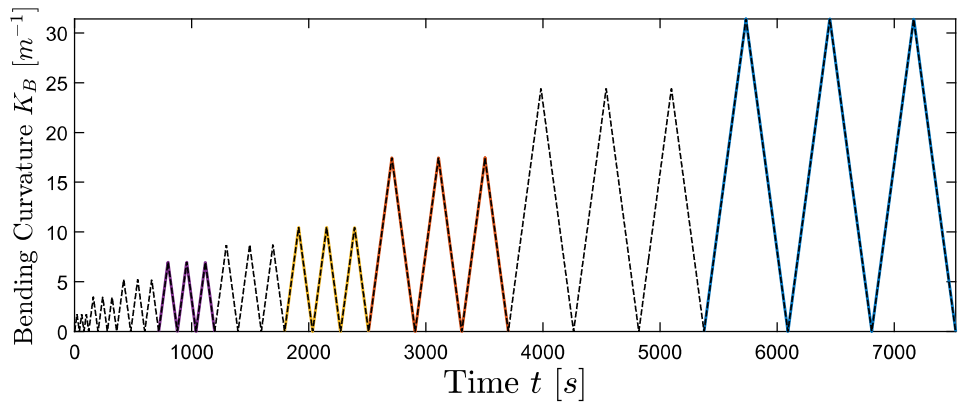

**Fig. 7** Input given as bending curvature vs. time. The purple, yellow, orange and blue parts are the input of Figs. 6, 9, 10, 11, respectively. (Colour figure online)

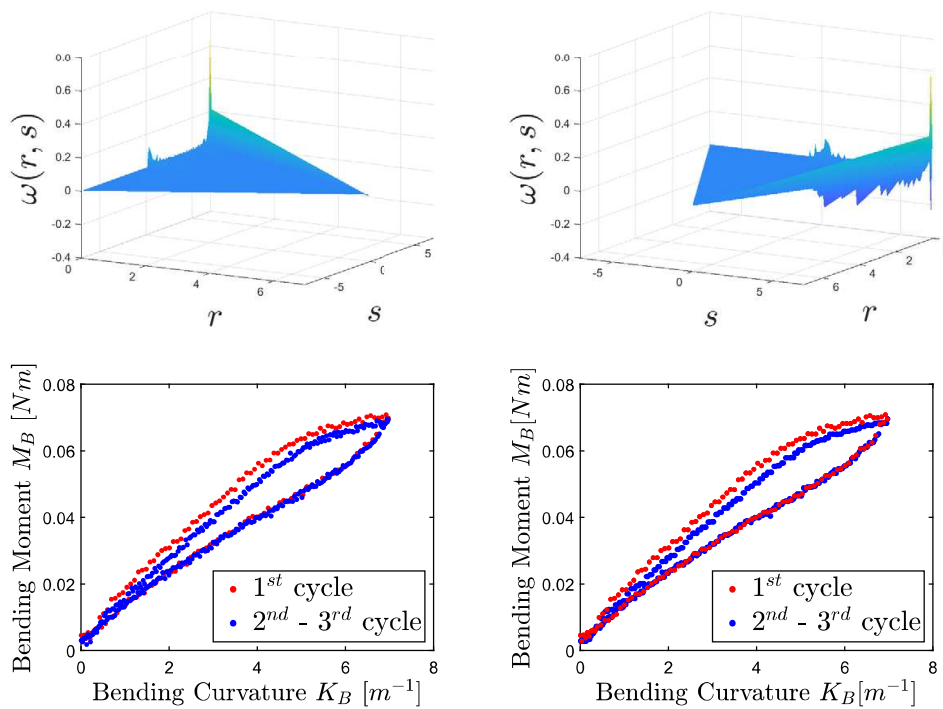

**Fig. 8** Results for cycles 10–12. *Top left, right:* estimated kernel function seen from two different angles. *Bottom left:* measured bending moment vs. bending curvature, starting from a piecewise linear input (bending curvature). *Bottom right:* estimated plot of bending moment vs. curvature. (Colour figure online)

curvature, and three cycles are executed on this level. This procedure is continued until a maximum bending curvature of 31.4 m$^{-1}$ is reached. This yields a total of 27 load cycles.

From these data, we extract those concerning loading cycles $10-12$ (purple) with maximal bending curvature $K_B^4 = 6.9$ m$^{-1}$, cycles $16-18$ (yellow) with $K_B^6 = 10.4$ m$^{-1}$, cycles $19-21$ (orange) $K_B^7 = 17.45$ m$^{-1}$ and cycles $25-27$ (blue) with $K_B^9 = 31.4$ m$^{-1}$. We treat these data as separate data sets and approximate separate kernel functions for each data set.

In Figs. 8, 9, 10, 11, we show on top the estimated kernel function seen from two different angles, on bottom left the measured values of the bending moment w.r.t bending curvature and on bottom right the estimated hysteresis diagram evaluated by means of the kernel function integrated over the suitable subset of the Preisach plane. One can notice that with this experiment, the difference in the hysteresis diagrams between the first cycle and the follow-ing ones is less visible, probably due to the fact that the cable is bent many times on one bending curvature, adding damage incrementally from one curvature level to the next. In the experiment shown in Fig. 2 (*right*), the whole damage accumulation occurs during the first load cycle.

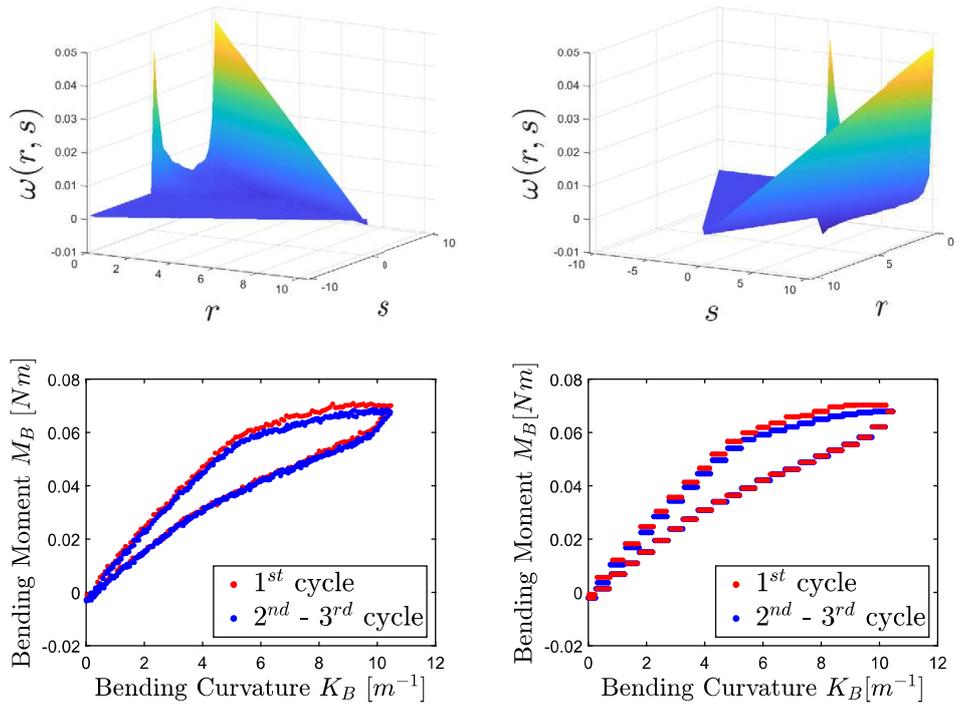

**Fig. 9** Results for cycles 16–18. *Top left, right:* estimated kernel function seen from two different angles. *Bottom left:* measured bending moment vs. bending curvature, starting from a piecewise linear input (bending curvature). *Bottom right:* estimated plot of bending moment vs. curvature. (Colour figure online)

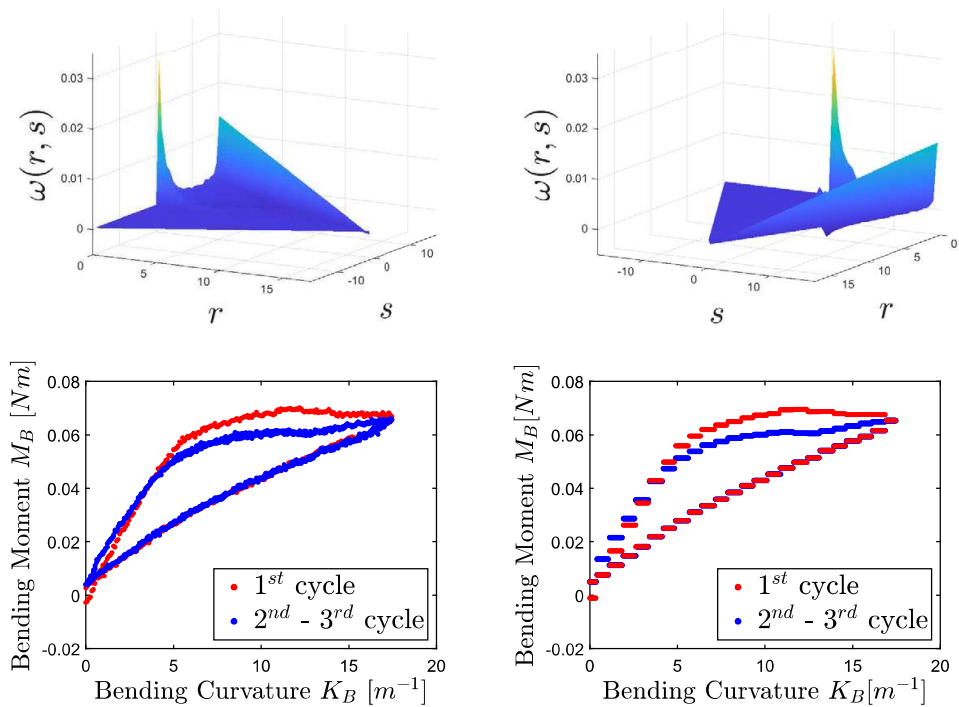

**Fig. 10** Results for cycles 19–21. *Top left, right:* estimated kernel function seen from two different angles. *Bottom left:* measured bending moment vs. bending curvature, starting from a piecewise linear input (bending curvature). *Bottom right:* estimated plot of bending moment vs. curvature. (Colour figure online)

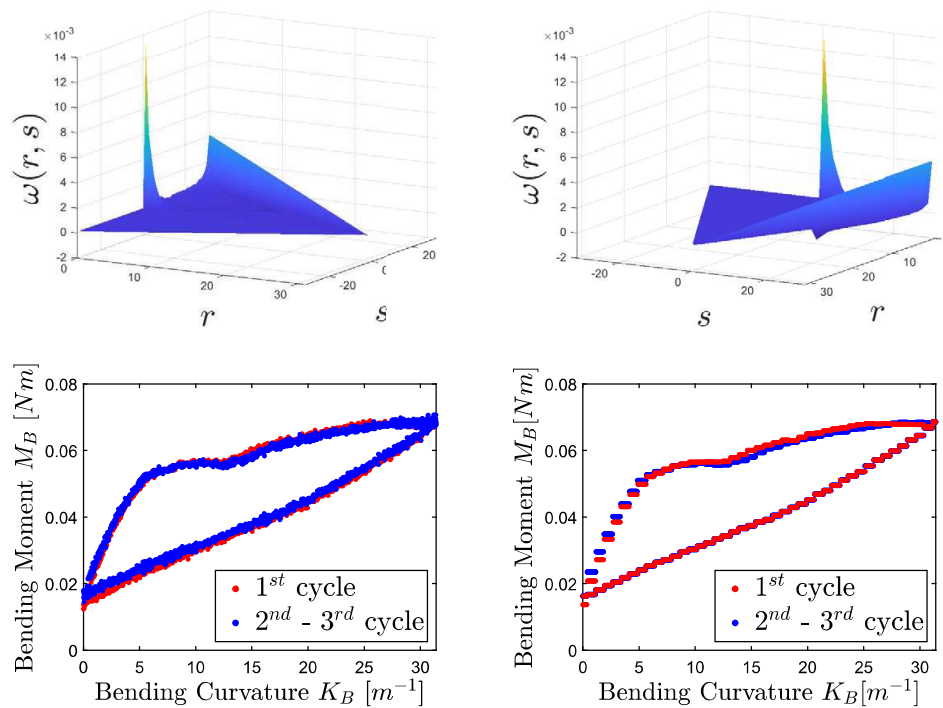

**Fig. 11** Results for cycles 25–27. *Top left, right:* estimated kernel function seen from two different angles. *Bottom left:* measured bending moment vs. bending curvature, starting from a piecewise linear input (bending curvature). *Bottom right:* estimated plot of bending moment vs. curvature. (Colour figure online)

At first, one can notice how the Preisach operator is able to reproduce very different hysteresis behaviours, either with noticeable difference between first and following loading cycles (Figs. 6, 8, 9, 10) and with static hysteresis (Fig. 11). Observations can be made by comparing Figs. 6 and 11. Both cases deal with similar maximal values of bending curvature as input. In Fig. 6, the bending moment in the first cycle differs from the following ones and the identified kernel function is nonsmooth. In Fig. 11, however, the bending moment shows a static hysteresis as there is no difference between the cycles and the kernel function appears smoother.

## 6 Summary, conclusions and outlook

In this work, we outlined our approach to utilise Cosserat rods as grey box models for cable simulation. The central element of such models, and a nonstandard as well as novel contribution compared to the usual applications of nonlinear rod models in structural mechanics and flexible multibody dynamics, are data based inelastic constitutive models formulated in terms of hysteresis operators adapted to capture the deformation behaviour of composite cables.

For this purpose, we introduced Preisach operators defined as a superposition of relay operators, explained an operative approach to compute the Preisach kernel function starting from experimental data which display a hysteresis behaviour and approximated such a function for different experimental data sets. We demonstrated that Preisach operators are a very powerful and versatile tool to describe inelastic deformations of electric cables, including open hysteresis loops arising from cyclic bending experiments. The data based constitutive model captures different hysteresis cycles very well and is relatively easy to implement. However, the interpretation and the comparison of kernel functions in different cases appears to be a nontrivial task, even when observing similar phenomena. The possible correlation between the properties of the kernel function and the physical phenomena should be the topic of further investigation.

Further steps in our future work will address the construction of a numerical procedure to solve the static equilibrium equations for a Cosserat rod model in the special case of plane quasistatic finite bending deformations, in the direction towards which [15] have already investigated the case of plane bending of an Euler Bernoulli beam in a geometrically linear context. As inelastic behaviour resulting in hysteretic response is also observed in cyclic twisting experiments of cables, we expect that a corresponding scalar constitutive model for torsional deformations can be identified in a similar manner as in the case of plane bending.

The simulation of spatial deformations of cables will require a combination of an inelastic constitutive model of vector type as formulated in Eq. (2) that either combines scalar hysteresis operators for both bending and twisting or handles such combined load cases in a coupled form. A related complication that might occur for composite cables due to their internal structure is a coupling of torsional behaviour with the longitudinal component of the sectional force. This is subject of current experimental investigations. However, we expect that identifying coupling effects between bending and torsion will be a rather difficult task.

**Acknowledgements** We thank Klaus Kuhnen and Pavel Krejčí for fruitful in depth discussions on the mathematical theory and implementation of hysteresis operators.

**Funding** Open Access funding enabled and organized by Projekt DEAL. This project has received funding from the European Union's Horizon 2020 research and innovation programme under the Marie Skłodowska-Curie grant agreement No 860124. This publication reflects only the authors' view, and the Research Executive Agency is not responsible for any use that may be made of the information it contains.

## Declarations

**Competing Interests** The authors declare no competing interests.